\documentclass[aps,pre,twocolumn,showpacs]{revtex4}
\usepackage{amsfonts}
\usepackage{amsmath}
\usepackage{amssymb}
\usepackage{graphicx}

\def\kbar{\protect\@kbar}
\def\@kbar{\relax \bgroup
\def\@tempa{\hbox{\raise.73\ht0
\hbox to0pt{\kern.25\wd0\vrule width.5\wd0 height.1pt
depth.1pt\hss}\box0}}\mathchoice{\setbox0\hbox{$\displaystyle
k$}\@tempa}{\setbox0\hbox{$\textstyle
k$}\@tempa}{\setbox0\hbox{$\scriptstyle
k$}\@tempa}{\setbox0\hbox{$\scriptscriptstyle k$}\@tempa}\egroup}

\begin{document}

\title{\textbf{Quantum Properties of Double Kicked Systems with Classical
Translational Invariance in Momentum}}
\author{Itzhack Dana}
\affiliation{Minerva Center and Department of Physics, Bar-Ilan University, Ramat-Gan
52900, Israel}

\begin{abstract}
Double kicked rotors (DKRs) appear to be the simplest nonintegrable
Hamiltonian systems featuring classical translational symmetry in phase
space (i.e., in angular momentum) for an \emph{infinite} set of values (the rational ones) of a parameter $\eta$. The experimental realization of
quantum DKRs by atom-optics methods motivates the study of the double kicked
particle (DKP). The latter reduces, at any fixed value of the conserved
quasimomentum $\beta\hbar$, to a generalized DKR, the \textquotedblleft 
$\beta $-DKR\textquotedblright . We determine general quantum properties of 
$\beta $-DKRs and DKPs for arbitrary rational $\eta $. The quasienergy
problem of $\beta $-DKRs is shown to be equivalent to the energy eigenvalue
problem of a finite strip of coupled lattice chains. Exact connections are then obtained between quasienergy spectra of $\beta $-DKRs for all $\beta $ in a generically infinite set. The general conditions of quantum resonance for $\beta $-DKRs are shown to be the simultaneous rationality of $\eta $, $\beta$, and a scaled Planck constant $\hbar _{\mathrm{S}}$. For rational $\hbar _{\mathrm{S}}$ and generic values of $\beta $, the quasienergy spectrum is found to have a staggered-ladder structure. Other spectral structures, resembling Hofstadter butterflies, are also found. Finally, we show the existence of particular DKP wave-packets whose quantum dynamics is \emph{free}, i.e., the evolution frequencies of expectation values in these wave-packets are independent of the nonintegrability. All the results for rational $\hbar _{\mathrm{S}}$ exhibit unique number-theoretical features involving $\eta $, $\hbar _{\mathrm{S}}$, and $\beta $.
\end{abstract}

\pacs{05.45.Ac, 03.65.Ca, 05.60.Gg, 03.75.Be}
\maketitle

\begin{center}
\textbf{I. INTRODUCTION}
\end{center}

Classical or quantum translational symmetry in a phase space is exhibited by
paradigmatic and realistic nonintegrable Hamiltonian systems and is useful
for their theoretical analysis \cite{bc,cm,krw,ikh,ihkm,d,d1,wm,wm1,gmz,cr1,cr2,cr3,cr4,dr,dd,gpf,fmi,fgr,wgf,d2,d3,khm,gkp,gkpr,d4,d5,dfw,dd1,lw,d6,kgp,d7}. It is also responsible for a variety of interesting classical and quantum phenomena. A well known example are classical transporting islands \cite{krw,ikh,ihkm,d,d1,wm1,gmz,cr1}, leading to chaotic-transport effects such as superdiffusion \cite{ikh,ihkm,d,d1,wm1,gmz} and Hamiltonian ratchets \cite{cr1,cr2,cr3,cr4}. These effects are most significant when they take place in the momentum direction due, e.g., to accelerator-mode islands. Quantum translational invariance in momentum leads to quantum resonances (ballistic quantum motion) \cite{gpf,fmi,fgr,wgf,d2,d3,dd1,eqr}. These and related phenomena, such as quantum accelerator modes \cite{fgr,qam} and quantum-resonance ratchets \cite{lw,d6,kgp,d7,shsn,drrts}, have been experimentally observed \cite{eqr,qam,shsn,drrts}.

Phase-space translational symmetry in classical systems may exist only for
special values of a control parameter. A well-known example is the Zaslavsky
web map \cite{wm}, for which these special values form a finite small set 
\cite{note}. The simplest systems featuring an \emph{infinite} dense set of
such parameter values are apparently the double kicked rotors (DKRs) \cite{cr2,cr3,des,1,2,3,4,5,6,7,8,9,10,11,12,13}, which has been considerably
studied during the last decade and appear to exhibit a rich variety of phenomena. A most general version of the DKR is
described by the time-periodic Hamiltonian 
\begin{eqnarray}
H_{\mathrm{DKR}} &=&\frac{L^{2}}{2}+KV(\theta )\sum_{t=-\infty }^{\infty
}\delta (t^{\prime }-t)  \notag \\
&+&\tilde{K}\tilde{V}(\theta )\sum_{t=-\infty }^{\infty }\delta (t^{\prime
}-\eta -t),  \label{HDKR}
\end{eqnarray}
where $(\theta ,L)$ are angle and angular momentum, $K$ and $\tilde{K}$ are
nonintegrability parameters, $V(\theta )$ and $\tilde{V}(\theta )$ are
general $2\pi $-periodic potentials, $t^{\prime }$ is the usual (continuous)
time, and $t^{\prime }=0,\eta $ ($0\leq \eta <1$) are the two kicking times
in one time period. The latter and the inertia moment of the rotor are
assumed to be $1$ in suitably chosen units. The DKR is a special case of the
modulated kicked rotor (or multi-kicked rotor) introduced in Ref. 
\cite{des}. As one can easily show (see Sec. II), the system (\ref{HDKR}) exhibits classical translational symmetry in phase space, namely in angular momentum $L$, for all rational values of $\eta $. Several studies have shown the
sensitive dependence of the classical phase space and chaotic transport on 
$\eta$ \cite{cr2,cr3}. Among interesting known results for the quantum DKR, we mention here the ones obtained for both rational and irrational $\eta $ in the
case of cosine potentials and $\hbar =4\pi$ \cite{6,7,8,9,10,11,12,13}, where 
$\hbar $ is the dimensionless Planck constant. In this case, the quasienergy (QE) spectrum as function of $\eta$ exhibits an approximate Hofstadter-butterfly \cite{gkp,hb} structure (see also Secs. IIIE and V).

The experimental realization of quantum DKRs by atom-optics methods
motivates the study of the double kicked particle (DKP), i.e., the system (\ref{HDKR}) with $(\theta ,L)$ replaced by scaled position and momentum
variables $(x,p)$. In such a realization, the particles (atoms) are
periodically kicked by two spatially periodic optical potentials $V(x)$ and $\tilde{V}(x)$ [Eq. (\ref{HDKR}) with $\theta \rightarrow x$]. As it is well
known \cite{fgr,wgf,d2,d6}, spatial periodicity leads to the conservation of a quasimomentum $\beta \hbar $, $0\leq \beta <1$. This allows to get exact and simple relations between general quantum kicked particles and quantum kicked rotors: At any fixed value of $\beta $, the DKP reduces to a generalized DKR, the \textquotedblleft $\beta $-DKR\textquotedblright , see Sec. IIIA.

In this paper, we determine general quantum properties of $\beta $-DKRs and
DKPs under conditions of classical translational invariance in angular
momentum, i.e., arbitrary rational $\eta $. The ordinary quantum DKR
investigated in previous works corresponds to $\beta =0$. Effects of nonzero
values of $\beta $ were considered in Refs. \cite{10,11}, but only in the case
of $\hbar =4\pi $ and using approximate approaches. Here, in
contrast, we study general values of $\beta $ and $\hbar $ on an exact,
first-principles basis.

The organization and contents of this paper are as follows. In Sec. II, we
consider some relevant implications of momentum translational invariance
(rational $\eta $) for classical DKRs; here and from now on, the term
\textquotedblleft momentum\textquotedblright\ means either angular momentum
or linear momentum, depending on whether DKRs or DKPs are considered. In
Sec. IIIA, we review the concept of quasimomentum $\beta \hbar $ in the
context of the quantum DKP and present the basic relation between the DKP and the $\beta $-DKRs. In the rest of Sec. III, we obtain several results concerning $\beta $-DKRs for arbitrary rational $\eta $. In Sec. IIIB, we show that the QE
problem for $\beta $-DKRs can be exactly mapped into the energy eigenvalue
problem of a finite strip of coupled lattice chains with translationally
invariant hopping constants; the chains are pseudorandom, quasiperiodic,
or periodic depending on the values of $\beta $ and of a scaled Planck
constant $\hbar _{\mathrm{S}}=\hbar /(2\pi )$. Using these chains,
we derive in Sec. IIIC exact connections between the QE spectra
of $\beta $-DKRs for all $\beta $ in a generically infinite set. In Sec.
IIID, we derive the general conditions for quantum resonance in 
$\beta $-DKRs; these are the simultaneous rationality of $\eta $, $\hbar _{\mathrm{S}}$, and $\beta $. In Sec. IIIE, we show that for rational $\hbar_{\mathrm{S}}$ but general $\beta $ the QE spectrum of $\beta $-DKRs has a staggered-ladder structure, i.e., it is a superposition of a finite number of ladder subspectra. We also find novel spectral structures for $\hbar\neq 4\pi$, resembling Hofstadter butterflies. In Sec. IV, we show that the quantum dynamics of particular DKP wave-packets is \emph{free}, i.e., the evolution frequencies of expectation values in these wave-packets are independent of the nonintegrability. All the results for rational $\hbar _{\mathrm{S}}$ exhibit unique number-theoretical features involving $\eta $, $\hbar _{\mathrm{S}}$, and $\beta $. A summary and conclusions are presented in Sec. V.

\begin{center}
\textbf{II. IMPLICATIONS OF MOMENTUM TRANSLATIONAL INVARIANCE FOR CLASSICAL
DKRs}
\end{center}

Defining the force functions $f(\theta )=-dV/d\theta $ and $\tilde{f}(\theta
)=-d\tilde{V}/d\theta $, the classical map for the Hamiltonian (\ref{HDKR})
in one time period is: 
\begin{equation}
\begin{array}{lll}
L_{t+1} & = & L_{t}+Kf(\theta _{t})+\tilde{K}\tilde{f}(\tilde{\theta}_{t}),
\\ 
&  &  \\ 
\theta _{t+1} & = & \tilde{\theta}_{t}+(1-\eta )L_{t+1}\ \text{\textrm{mod}}
(2\pi ),
\end{array}
\label{map}
\end{equation}
where $(\theta _{t},L_{t})$ denote the values of $(\theta ,L)$ at time 
$t^{\prime }=t-0$ and 
\begin{equation}
\tilde{\theta}_{t}=\theta _{t}+\eta L_{t}+K\eta f(\theta _{t}).  \label{tt}
\end{equation}
It is is easy to see that the map (\ref{map}) with (\ref{tt}) is invariant
under a translation $L^{\prime }$ in momentum $L$ only if $\eta L^{\prime
}=2\pi a$ and $\eta L^{\prime }+(1-\eta )L^{\prime }=L^{\prime }=2\pi c$,
where $a$ and $c$ are integers. Therefore, $\eta $ must be rational,
\begin{equation}
\eta =\frac{a}{c},  \label{eta}
\end{equation}
where $a$ and $c$ can be assumed to be coprime integers. The map (\ref{map})
may then be defined on the torus $\mathbb{T}_{\mathrm{c}}^{2}:0\leq \theta
<2\pi $, $0\leq L<2\pi c$, by taking $L_{t+1}$ modulo $2\pi c$. The 
$2\pi c$-periodicity in $L$ of some orbits of the map (\ref{map}) is illustrated in Fig. 1. 
\begin{figure}[tbp]
\includegraphics[width=9.0cm]{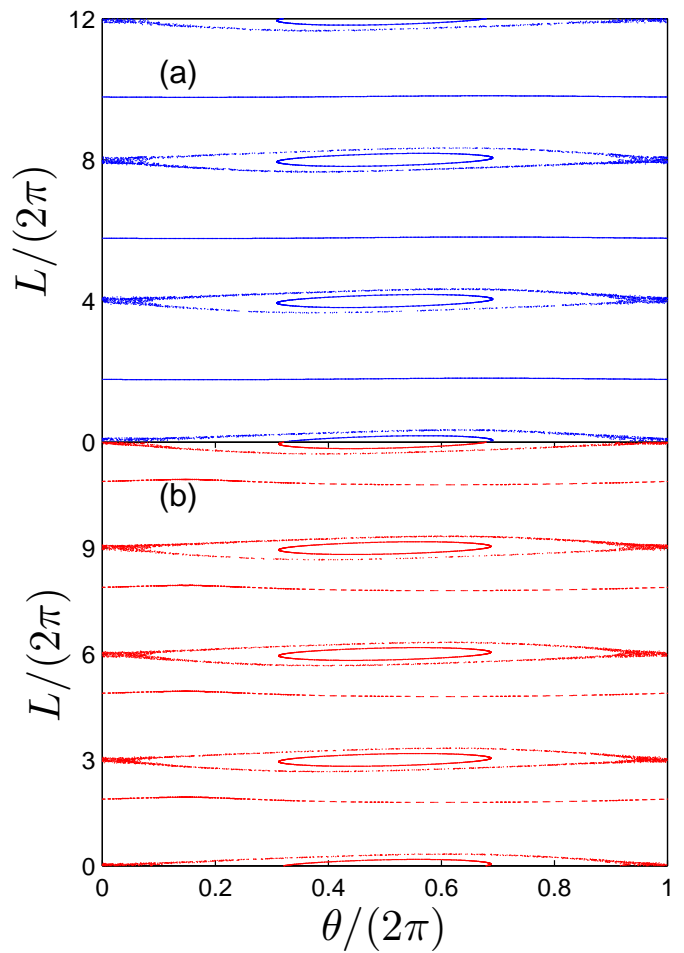}
\caption{(Color online) Some orbits of the DKR map (\ref{map}) for $
K=\tilde{K}=0.5$, $V(\theta )=\tilde{V}(\theta )=\cos (\theta )$, and (a) $\eta =1/4$; (b) $\eta =1/3$. The periodicity in $L/(2\pi )$ with periods $c=4$ and $c=3$ in (a) and (b), respectively, is evident. The three basic orbits in the first unit cell, with $0\leq L/(2\pi )<c$, are a vibrational orbit, a chaotic orbit in the main stochastic layer, and a rotational orbit. Each of these orbits starts from the same initial conditions in (a) and (b).}
\label{fig1}
\end{figure}

We now consider some implications of the momentum translational invariance.
First, there may exist periodic accelerator modes satisfying $\theta
_{t+m}=\theta _{t}$, $\ L_{t+m}=L_{t}+2\pi cw$, where $m$ is the minimal
period and $w$ is a nonzero integer. This is because period-$m$ orbits of
the torus map [Eq. (\ref{map}) with $L_{t+1}$ taken modulo $2\pi c$],
satisfying $L_{t+m}=L_{t}$, may generally satisfy $L_{t+m}=L_{t}+2\pi cw$
when considered on the original (cylindrical) phase space. Second, consider
the mean energy of a closed ($w=0$) period-$m$ orbit for the Hamiltonian 
(\ref{HDKR}): 
\begin{equation}
\bar{E}=\frac{1}{m}\int_{-0}^{m-0}dt^{\prime }H_{\text{\textrm{DKR}}
}(t^{\prime })=\frac{1}{m}\sum_{t=0}^{m-1}\bar{E}_{t},  \label{mE}
\end{equation}
where $\bar{E}_{t}$ is the mean energy during the $t$th time period,
\begin{eqnarray}
\bar{E}_{t} &=&\int_{t-0}^{t+1-0}dt^{\prime }H_{\text{\textrm{DKR}}
}(t^{\prime }) = \eta \frac{\left[ L_{t}+Kf(\theta _{t})\right] ^{2}}{2} 
\notag \\
&+&(1-\eta )\frac{L_{t+1}^{2}}{2}+KV(\theta _{t})+\tilde{K}\tilde{f}(\tilde{
\theta}_{t}),  \label{Es}
\end{eqnarray}
with $\tilde{\theta}_{t}$ defined by Eq. (\ref{tt}). Given a period-$m$
orbit $(\theta _{t},L_{t})$ with mean energy (\ref{mE}), its translation 
$(\theta _{t},L_{t}+2\pi cj)$ in $L$ by $2\pi cj$ ($j$ integer) is also a
period-$m$ orbit. Using Eqs. (\ref{mE}) and (\ref{Es}), we find that the
mean energy of the latter orbit is
\begin{equation}
\bar{E}^{(j)}=\bar{E}+2\pi cj\bar{L}+2\pi ^{2}c^{2}j^{2},  \label{mEj}
\end{equation}
where $\bar{L}$ is the mean momentum of the original orbit: 
\begin{equation}
\bar{L}=\frac{1}{m}\sum_{t=0}^{m-1}\left[ \eta \left( L_{t}+Kf(\theta
_{t})\right) +(1-\eta )L_{t+1}\right] .  \label{mL}
\end{equation}
The mean momentum of the translated orbit is: 
\begin{equation}
\bar{L}^{(j)}=\bar{L}+2\pi cj.  \label{mLj}
\end{equation}

\begin{center}
\textbf{III. QUANTUM PROPERTIES OF DKP AT FIXED QUASIMOMENTUM}

\textbf{A. Quantum DKP, Quasimomentum, and }$\boldsymbol{\beta}$\textbf{-DKRs}
\end{center}

For the quantum DKP, described by the Hamiltonian (\ref{HDKR}) with $\theta
\rightarrow \hat{x}$ and $L\rightarrow \hat{p}$ (carets indicate operators),
the one-period evolution operator from $t^{\prime }=t-0$ to $t^{\prime
}=t+1-0$ is given by
\begin{eqnarray}
\hat{U}(\hat{x},\hat{p}) &=&\exp \left[ -i\left( 1-\eta \right)\frac{\hat{p}
^{2}}{2\hbar }\right] \exp \left[ -i\frac{\tilde{K}\tilde{V}(\hat{x})}{\hbar 
}\right]  \notag \\
&\times &\exp \left( -i\eta \frac{\hat{p}^{2}}{2\hbar }\right) \exp \left[ -i\frac{KV(\hat{x})}{\hbar }\right] .  \label{U}
\end{eqnarray}
Here $[\hat{x},\hat{p}]=i\hbar $, where $\hbar $ is the dimensionless Planck
constant. The QE eigenvalue problem for (\ref{U}) is
\begin{equation}
\hat{U}(\hat{x},\hat{p})\Psi _{\omega }(x)=\exp (-i\omega )\Psi _{\omega
}(x),  \label{ep}
\end{equation}
where $\omega $ is the QE, $0\leq \omega <2\pi $. Because of the 
$2\pi$-periodicity of (\ref{U}) in $\hat{x}$, the QE state $\Psi _{\omega }(x)$ can be chosen to have the Bloch form:
\begin{equation}
\Psi _{\omega }(x)=\exp (i\beta x)u_{\beta ,\omega }(x),  \label{qes}
\end{equation}
where $\beta $, $0\leq \beta <1$, gives the quasimomentum $\beta \hbar $ and 
$u_{\beta ,\omega }(x)$ is $2\pi $-periodic in $x$. Inserting (\ref{qes})
into (\ref{ep}) and using the fact that $\exp (i\beta \hat{x})=\exp (-\beta
\hbar d/dp)$ is a translation of $\hat{p}$ by $-\beta \hbar $, we find that $u_{\beta ,\omega }(x)$ satisfies the eigenvalue equation $\hat{U}(\hat{x},
\hat{p}+\beta \hbar )u_{\beta ,\omega }(x)=\exp (-i\omega )u_{\beta ,\omega
}(x)$. Due to the $2\pi$-periodicity of $u_{\beta ,\omega }(x)$, one can
interpret $x$ in the latter equation as an angle $\theta $ and $\hat{p}$ in $\hat{U}(\hat{x},\hat{p}+\beta \hbar )$ as an angular-momentum operator $\hat{L}$, since $\hat{p}/\hbar $ can only have integer eigenvalues $n$ when
acting on $2\pi $-periodic functions $u_{\beta ,\omega }(x)$. One can then
write the eigenvalue equation for $u_{\beta ,\omega }(x)$ as 
\begin{equation}
\hat{U}(\theta ,\hat{L}+\beta \hbar )u_{\beta ,\omega }(\theta )=\exp
(-i\omega )u_{\beta ,\omega }(\theta ).  \label{bep}
\end{equation}
In this way, the eigenvalue problem (\ref{ep}) at each fixed value of $\beta 
$ reduces to that of a \textquotedblleft $\beta $-DKR\textquotedblright\
with one-period evolution operator $\hat{U}(\theta ,\hat{L}+\beta \hbar )$.
The expression $\hat{L}+\beta \hbar $ corresponds to the decomposition of
the momentum eigenvalues $p=n\hbar +\beta \hbar $ into the eigenvalues 
$n\hbar $ of $\hat{L}$ and the \textquotedblleft
fractional\textquotedblright\ part $\beta \hbar $, the fixed quasimomentum.

\begin{center}
\textbf{B. Strip of Coupled Lattice Chains for }$\boldsymbol{\beta }$\textbf{-DKRs}
\end{center}

It is well known \cite{gpf} that the QE problem of the usual quantum kicked
rotor can be mapped into the energy-eigenvalue problem of a lattice chain
which is generically pseudorandom and whose eigenstates exhibit exponential
Anderson-like localization in angular-momentum space. The generalization of
this mapping to modulated or multi-kicked rotors \cite{des}, with $M$ kicks
of different strengths within one time period $T$, can be performed in the
simplest way when the $M$ kicks are equidistant in time ($\Delta T=T/M$);
the mapping then leads to a strip of $M$ coupled lattice chains having the advantage
that the hopping constants are translationally invariant.

In the case of the $\beta $-DKRs (see Sec. IIIA), the $M=2$ kicks in one
time period are generally not equidistant in time. However, for rational 
$\eta =a/c$ [Eq. (\ref{eta})], one can divide the time period $T=1$ into $c$
equal time intervals ($\Delta T=1/c$), labeled by $k=0,\dots ,c-1$, so that
kicks take place only at the beginning of the two intervals $k=0$ and $k=a$.
The QE problem (\ref{bep}) for the $\beta $-DKR can then be mapped into a relatively simple strip of $c$ coupled lattice chains as follows. We denote by $u_{k}^{\pm }(\theta )$ the QE eigenstate of the $\beta $-DKR at times $t^{\prime }=k/c\pm 0$ and by $u_{k,n}^{\pm }$ ($n$ integer) the angular-momentum representation of 
$u_{k}^{\pm }(\theta )$. The free quantum motion of the $\beta $-DKR within
each of the $c$ intervals is expressed by
\begin{gather}
u_{k,n}^{-}=\exp \left[ -i\hbar \left( n+\beta \right) ^{2}/(2c)\right]
u_{k-1,n}^{+}\ \ \ (0<k\leq c-1),  \label{fqm1} \\
\exp \left[ -i\hbar \left( n+\beta \right) ^{2}/(2c)\right]
u_{c-1,n}^{+}=\exp (-i\omega )u_{0,n}^{-},  \label{fqm2}
\end{gather}
where Eq. (\ref{fqm2}) follows from the fact that $u_{0,n}^{-}$ is a QE
eigenstate. Next, we define
\begin{equation}
u_{k}(\theta )=\exp \left( i\frac{k\omega }{c}\right) \frac{u_{k}^{-}(\theta
)+u_{k}^{+}(\theta )}{2}  \label{ut}
\end{equation}
and write the contributions of kicks to the evolution operator in the Cayley
form
\begin{equation}
\exp \left[ -iV_{k}(\theta )/\hbar \right] =\frac{1+iW_{k}(\theta )}{
1-iW_{k}(\theta )},  \label{cf}
\end{equation}
where $V_{k}(\theta )=0$, except for $V_{0}(\theta )=KV(\theta )$ and 
$V_{a}(\theta )=\tilde{K}\tilde{V}(\theta )$, and $W_{k}(\theta )=-\tan \left[ V_{k}(\theta )/(2\hbar )\right] $ from inversion of Eq. (\ref{cf}). Denoting by $u_{k,n}$ and $W_{k,n}$ the angular-momentum representations of $u_{k}(\theta )$ and $W_{k}(\theta )$, respectively, we get from Eqs. 
(\ref{fqm1})-(\ref{cf}) after simple algebra a system of $c$ equations:
\begin{align}
& u_{k+1,n}-i\sum_{r=-\infty }^{\infty }W_{k+1,n-r}u_{k+1,r}  \notag \\
& =e^{i\left[ \omega -\hbar \left( n+\beta \right) ^{2}/2)\right] /c}\left(
u_{k,n}+i\sum_{r=-\infty }^{\infty }W_{k,n-r}u_{k,r}\right) ,  \label{se}
\end{align}
$k=0,\dots ,c-1$, and $u_{c,n}=u_{0,n}$. The Fourier transforms of $u_{k,n}$ and $W_{k,n}$ in the discrete variable $k$ are given by
\begin{equation}
\bar{u}_{s,n}=\frac{1}{c}\sum_{k=0}^{c-1}u_{k,n}\exp \left( 2\pi
iks/c\right) ,  \label{usn}
\end{equation}
\begin{eqnarray}
\bar{W}_{s,n} &=&\frac{1}{c}\sum_{k=0}^{c-1}W_{k,n}\exp \left( 2\pi
iks/c\right)  \notag \\
&=&\frac{W_{0,n}+W_{a,n}\exp \left( 2\pi i\eta s\right) }{c}  \label{Wsn}
\end{eqnarray}
($s=0,\dots ,c-1$); notice the simple expression (\ref{Wsn}) for $\bar{W}_{s,n}$. Equations (\ref{se}) can then be easily written in a compact form in terms of (\ref{usn}) and (\ref{Wsn}), using the inverse Fourier-transform relations:
\begin{equation}
T_{n}^{(s)}\bar{u}_{s,n}+\sum_{r=-\infty }^{\infty }\sum_{s^{\prime }=0}^{c-1 }{}^{\prime }\ \bar{W}_{s-s^{\prime },n-r}\bar{u}_{s^{\prime },r}=E\bar{u}_{s,n},  \label{mcs}
\end{equation}
where $E=-\bar{W}_{0,0}=-(W_{0,0}+W_{a,0})/c$,
\begin{equation}
T_{n}^{(s)}=\tan \left[ \frac{\omega +2\pi s-\pi \hbar _{\mathrm{S}}(n+\beta
)^{2}}{2c}\right]  \label{Tsn}
\end{equation}
with $\hbar _{\mathrm{S}}=\hbar /(2\pi )$, and the prime after the sums means that the sums do not include the element with $r=n$ and $s^{\prime}=s$. Equations (\ref{mcs}) describe a strip of $c$ coupled lattice chains $s=0,\dots ,c-1$ with on-site potential $T_{n}^{(s)}$ and translationally invariant hopping constants $\bar{W}_{s-s^{\prime },n-r}$ from site $(s,n)$ to site $(s^{\prime},r)$. The eigenvalue problem (\ref{mcs}) is fully equivalent to the QE problem (\ref{bep}) for the $\beta $-DKR: The QE spectrum for given $\beta $ consists of all the values of $\omega $ in (\ref{Tsn}) such that $E=-\bar{W}_{0,0}$ is an \textquotedblleft energy\textquotedblright eigenvalue of Eqs. (\ref{mcs}). 
If $\bar{W}_{s-s^{\prime},n-r}$ is of sufficiently short range in $n-r$, each lattice chain is a tight-binding one. For irrational $\hbar _{\mathrm{S}}$, the potential $T_{n}^{(s)}$ in Eq. (\ref{Tsn}) is a pseudorandom function of $n$ and one expects exponential Anderson-like localization of the eigenstates $\bar{u}_{s,n}$ in $n$ and a pure point QE spectrum $\omega $, as for the usual kicked rotor \cite{gpf}. For rational $\hbar _{\mathrm{S}}$ and irrational $\beta $, $T_{n}^{(s)}$ is essentially quasiperiodic in $n$. Exact \cite{d2} and numerical \cite{d3} results for the ordinary kicked particle ($\eta =0$) indicate that also in this case the QE eigenstates should be localized in $n$ and the QE spectrum is pure point (see also end of Sec. IIIE). Finally, when both $\hbar _{\mathrm{S}}$ and $\beta $ are rational, $T_{n}^{(s)}$ is periodic in $n$, so that one expects a band continuous QE spectrum and extended QE eigenstates in $n$. In fact, this is the case of quantum resonance, to be considered in more detail in Sec. IIID.

\begin{center}
\textbf{C. Exact Connections between QE Spectra at Different Quasimomenta}
\end{center}

We now show that the results in Sec. IIIB lead to exact connections between
the QE\ spectra of $\beta $-DKRs for a generally infinite set of $\beta $
values. As we have seen in Sec. II, for rational $\eta =a/c$ the classical
map (\ref{map}) is invariant under translations by $2\pi cj$ ($j$ integer)
in $L$. For the quantum DKP, these translations correspond to the operators $\hat{D}_{2\pi cj}=\exp (2\pi icj\hat{x}/\hbar )=\exp (-2\pi cjd/dp)$. By
applying $\hat{D}_{j}$ to the Bloch state (\ref{qes}) of the DKP, we get:
\begin{eqnarray}
\hat{D}_{2\pi cj}\Psi _{\omega }(x) &=&\exp \left[ i(\beta +cj/\hbar _{
\mathrm{S}})x\right] u_{\beta ,\omega }(x)  \notag \\
&=&\exp (i\beta ^{(j)}x)u_{\beta ^{(j)},\omega ^{(j)}}(x),  \label{qesj}
\end{eqnarray}
where $\hbar _{\mathrm{S}}=\hbar /(2\pi )$, and
\begin{eqnarray}
\beta ^{(j)} &=&\beta +cj/\hbar _{\mathrm{S}}\ \ \text{\textrm{mod}}(1),
\label{bj} \\
u_{\beta ^{(j)},\omega ^{(j)}}(x) &=&\exp (in^{(j)}x)u_{\beta ,\omega }(x).
\label{ubj}
\end{eqnarray}
Here $n^{(j)}$ is the integer part of $\beta +cj/\hbar _{\mathrm{S}}$.
Equations (\ref{qesj})-(\ref{ubj}) mean that $\hat{D}_{2\pi cj}$ transforms
the QE state $u_{\beta ,\omega }(\theta )$ of a $\beta $-DKR into the QE
state $u_{\beta ^{(j)},\omega ^{(j)}}(\theta )$ of the $\beta ^{(j)}$-DKR.
The relation between the QEs $\omega $ and $\omega ^{(j)}$ of the two
states can be found using Eqs. (\ref{mcs}) and (\ref{Tsn}) for the
lattice chains. We easily see that the latter equations are invariant
under the simultaneous transformations $\beta \rightarrow \beta ^{(j)}$, $n\rightarrow n+n^{(j)}$, $r\rightarrow r+n^{(j)}$, and $\omega \rightarrow
\omega ^{(j)}$, where
\begin{equation}
\omega ^{(j)}=\omega +2\pi cj\beta +2\pi ^{2}c^{2}j^{2}/\hbar \ \ \text{
\textrm{mod}}(2\pi ).  \label{qej}
\end{equation}

Relation (\ref{qej}) now provides simple connections between the QE spectra
of the $\beta ^{(j)}$-DKRs for all $\beta ^{(j)}$ in Eq. (\ref{bj}): These
spectra are just shifted relative to each other by constant amounts. For
example, the spectrum for $\beta =\beta ^{(j)}$ is shifted relative to that
for $\beta =\beta ^{(0)}$ by $2\pi cj\beta ^{(0)}+2\pi ^{2}c^{2}j^{2}/\hbar
\ $\textrm{mod}$(2\pi )$. The spectral shifts and $\beta ^{(j)}$ are
independent of the system details. We remark that for generic, irrational
values of $\hbar _{\mathrm{S}}$ the set (\ref{bj}) covers \emph{densely} the
entire $\beta $ range $[0,1)$. For rational $\hbar _{\mathrm{S}}=l/q$ ($l$
and $q$ are coprime integers), let us write $c/l=c^{\prime }/l^{\prime }$,
where $c^{\prime }$ and $l^{\prime }$ are coprime; the set (\ref{bj}) then
contains only $l^{\prime }$ elements and the spectral shifts above are given
by $2\pi cj\beta ^{(0)}+\pi cc^{\prime }j^{2}q/l^{\prime }$. The integers 
$c^{\prime }$ and $l^{\prime }$ in $c/l=c^{\prime }/l^{\prime }$ depend
erratically on both the rational value of $\eta $ and of $\hbar _{\mathrm{S}}$. This number-theoretical feature will appear again below.

It is instructive to compare relation (\ref{qej}) with the classical one 
(\ref{mEj}). These relations coincide under the correspondences $\hbar\omega
^{(j)}\rightarrow \bar{E}^{(j)}$, $\hbar \omega \rightarrow \bar{E}$, and 
$\beta \hbar \rightarrow \bar{L}$. Similarly, the quasimomenta $\beta \hbar
^{(j)}$, with $\beta ^{(j)}$ given by Eq. (\ref{bj}), are analogous to the
translated classical angular momenta (\ref{mLj}). However, while $\bar{L}$
in Eq. (\ref{mLj}) depends on the specific kicking forces [see Eq. 
(\ref{mL})], this is not the case for $\beta$, which is a pure conserved quantum entity not depending on the system details.

\begin{center}
\textbf{D. General Quantum Resonances of }$\boldsymbol{\beta }$\textbf{-DKRs}
\end{center}

Quantum resonance (QR) in kicked-rotor systems is a ballistic (quadratic in
time) evolution of the expectation value of the kinetic energy due to a band
QE spectrum \cite{gpf,fmi,fgr,wgf,d2,d3,dd1,lw,d6}. This spectrum is a
consequence of quantum translational symmetry in $\hat{L}$. We derive here
the general conditions for this symmetry and for QR in $\beta $-DKRs. One condition turns out to be, as expected, the rationality (\ref{eta}) of $\eta $, i.e., classical
translational symmetry in $L$. The other conditions and their derivation
exhibit unique number-theoretical features.

A quantum translation in $\hat{L}$ is generally given by the operator $\hat{D}_{\bar{q}\hbar }=\exp (-i\bar{q}\theta )$, since $\hat{D}_{\bar{q}\hbar}
\hat{L}=(\hat{L}+\bar{q}\hbar )\hat{D}_{\bar{q}\hbar}$. Here $\bar{q}$ in $\exp
(-i\bar{q}\theta )$ must be integer since $\theta $ is an angle.
Translational invariance of a $\beta $-DKR in $\hat{L}$ means that $\left[ 
\hat{D}_{\bar{q}\hbar },\hat{U}(\theta ,\hat{L}+\beta \hbar )\right] =0$,
where $\hat{U}(\theta ,\hat{L}+\beta \hbar )$ is the one-period evolution
operator of a $\beta $-DKR defined by Eq. (\ref{U}). Then, as shown in
Appendix A, $\left[ \hat{D}_{\bar{q}\hbar },\hat{U}(\theta ,\hat{L}+\beta
\hbar )\right] =0$ implies that:
\begin{equation}
(1-\eta )\bar{c}=d,  \label{ceta}
\end{equation}
\begin{equation}
\bar{q}\hbar =2\pi \bar{c},  \label{hbar}
\end{equation}
\begin{equation}
\beta \bar{q}\hbar +\frac{\bar{q}^{2}\hbar }{2}=2\pi \bar{r},  \label{cb}
\end{equation}
where $\bar{c}$, $d$, and $\bar{r}$ are integers. From Eq. (\ref{ceta}) it follows that $\eta =\bar{a}/\bar{c}$, where $\bar{a}=\bar{c}-d$. Let the maximal
common factor of $(\bar{a},\bar{c})$ be $f_{\eta }$, so that $\bar{a}
=f_{\eta }a$ and $\bar{c}=f_{\eta }c$, where $(a,c)$ are coprime; then, 
$\eta =a/c$, which is condition (\ref{eta}). Next, we denote the maximal
common factor of $(\bar{c},\bar{q})$ by $f_{h}$, so that $\bar{c}=f_{h}l$
and $\bar{q}=f_{h}q$, where $(l,q)$ are coprime. Equation (\ref{hbar}) then
becomes
\begin{equation}
\hbar _{\mathrm{S}}=\frac{\hbar}{2\pi}=\frac{l}{q}.  \label{rh}
\end{equation}
Now, let $c/l=c^{\prime }/l^{\prime }$, where $(c^{\prime },l^{\prime })$
are coprime. Then, from $\bar{c}=f_{\eta }c=f_{h}l$, we get $f_{\eta
}c^{\prime }=f_{h}l^{\prime }$, implying that $f_{\eta }=gl^{\prime }$ and 
$f_{h}=gc^{\prime }$, where $g$ is some integer. Thus,
\begin{equation}
\bar{q}=f_{h}q=gc^{\prime }q.  \label{qb}
\end{equation}
Using Eqs. (\ref{rh}) and (\ref{qb}) in Eq. (\ref{cb}), we obtain:
\begin{equation}
\beta =\frac{\bar{r}}{gc^{\prime }l}-\frac{gc^{\prime }q}{2}\ \ \text{\textrm{mod}}(1).  \label{rb}
\end{equation}
For given $\eta =a/c$ and $\hbar $ [Eq. (\ref{rh})], any rational value 
$\beta _{\mathrm{r}}$ of $\beta $ can be expressed in the form (\ref{rb}) by
choosing $g$ as the smallest integer such that $gc^{\prime }l(\beta _{
\mathrm{r}}+gc^{\prime }q/2)$ is an integer $\bar{r}$. A simple and useful formula for the maximal value $g_{\rm max}$ of $g$ over all $\eta$ can be easily derived. Let us denote the denominator of $\beta _{\mathrm{r}}$ by $d_{\beta }$ and let $l/d_{\beta }=\bar{l}/d_{\beta }^{\prime }$, where $(\bar{l},d_{\beta }^{\prime })$ are coprime. We also define $\tau (lq)$ by: $\tau (lq)=1$ for $lq$ even and $\tau (lq)=2$ for $lq$ odd. Then
\begin{equation}
g_{\rm max}=\tau (lq)d_{\beta }^{\prime }. \label{gb}
\end{equation}
As shown by examples in Sec. IIIE, $g$ is equal to $g_{\rm max}$ for most values of $\eta$. The value of $g$ for any $\eta $ gives the minimal translation $\bar{q}\hbar =2\pi gc^{\prime }l=2\pi gl^{\prime }c$ in $\hat{L}$. The latter defines the \textquotedblleft quantum torus\textquotedblright\ $\mathbb{T}_{\mathrm{q}}^{2}:0\leq \theta <2\pi $, $0\leq L<2\pi gl^{\prime}c $, which is $gl^{\prime }$ larger than the classical torus $\mathbb{T}_{\mathrm{c}}^{2}$ in Sec. II. In summary, the general QR conditions for $\beta $-DKRs are arbitrary rational values of $\eta $, $\hbar _{\mathrm{S}}=\hbar /(2\pi )$, and $\beta $. These values determine $g$ as above. Then, $\bar{q}$ is found from Eq. (\ref{qb}) and $\mathbb{T}_{\mathrm{q}}^{2}$ is determined.

As shown in Appendix B, the QE spectrum of a $\beta $-DKR under QR conditions generally consists of $\bar{q}$ bands. QR will actually arise only if at least one band is not flat, i.e., has nonzero width. In the case that all bands are flat, QR is replaced by the opposite phenomenon of quantum antiresonance \cite{d2,des}, a bounded quantum motion.

\begin{center}
\textbf{E. Staggered-Ladder QE Spectra and Other Spectral Structures}
\end{center}

Having derived the general QR conditions for $\beta $-DKRs, including the
rationality of $\hbar _{\mathrm{S}}=\hbar /(2\pi )$, we now examine in more
detail the nature of the QE spectra for rational $\hbar _{\mathrm{S}}$ and arbitrary $\beta $, i.e., also in the non-QR case of
irrational $\beta $. Writing $c/l=c^{\prime }/l^{\prime }$ with $(c^{\prime
},l^{\prime })$ coprime integers as above, we see that the sequence  (\ref{bj}) is periodic with minimal period $l^{\prime }$: $\beta ^{(sl^{\prime })}=\beta ^{(0)}=\beta$, for all integers $s$. This means that the subsequence $\omega ^{(sl^{\prime })}$ of (\ref{qej}) is necessarily a QE subspectrum of the $\beta $-DKR. By simple algebra, using also $cl^{\prime }=c^{\prime }l$, we find that
\begin{eqnarray}
\omega ^{(sl^{\prime })} &=&\omega +2\pi sc^{\prime }l\beta +\pi
s^{2}c^{\prime 2}lq\ \ \text{\textrm{mod}}(2\pi )  \notag \\
&=&\omega +2\pi sc^{\prime }l\left( \beta +\frac{c^{\prime }q}{2}\right) \ \ 
\text{\textrm{mod}}(2\pi ).  \label{qerl}
\end{eqnarray}
Thus, the QE subspectrum (\ref{qerl}) is just a ladder, with spacing $\Delta
\omega =2\pi cl^{\prime }(\beta +c^{\prime }q/2)$, folded into the $[0,2\pi
) $ interval. We now show that the entire QE spectrum of the $\beta $-DKR is
a staggered ladder, i.e., the superposition of $c^{\prime }q$ ladders (\ref{qerl})
of bands or levels, corresponding to $c^{\prime }q$ independent values of 
$\omega $ in Eq. (\ref{qerl}). This was shown in Ref. \cite{d3}\ in the
special case of $\beta $-kicked-rotors with $\eta =0$ in Eq. (\ref{HDKR}),
so that $a=0$, $c=c^{\prime }=1$, and $l^{\prime }=l$.

Let us start with the QR case of rational $\beta $. The sequence (\ref{qerl}) is then periodic with minimal period $g$ ($\omega ^{(gl^{\prime })}=\omega
)$, where $g$ is the smallest integer such that $gc^{\prime }l(\beta
+c^{\prime }q/2)$ is an integer $\bar{r}$, obviously coprime to $g$; this gives
Eq. (\ref{rb}) \cite{note1}. Thus, the ladder subspectrum (\ref{qerl})
contains precisely $g$ levels. However, the full spectrum consists of $\bar{q}=gc^{\prime }q$ bands, see Sec. IIID and Appendix B. Therefore, this
spectrum must be a superposition of $c^{\prime }q$ ladders of bands, where
each ladder contains $g$ bands that arise from the broadening of the $g$
levels above. 

This staggered-ladder spectrum as function of $\eta$ is illustrated in Figs. 2(b), 2(c), 3(c), and 4(b) for different values of $\hbar$ and $\beta$, showing that the dominant value of $g$ over $\eta$ is given by $g_{\rm max}$ in Eq. (\ref{gb}). In Figures 2(b) and 2(c), both for $\hbar =4\pi$, the width of each of the $g$ stages of the ladder structure decreases as $\eta$ approaches $0$ or $0.5$. This is because for $\eta =0,0.5$ the DKR (\ref{HDKR}) reduces to an ordinary kicked rotor; it is known \cite{d2} that for $\hbar =2\pi l$, $\beta$ with $g>1$, and cosine potentials as in Fig. 2, a $\beta$-kicked-rotor exhibits quantum antiresonance (instead of QR) with a QE spectrum consisting of $g$ flat bands. Thus, the width of each stage in Figs. 2(b) and 2(c) must tend to zero as $\eta\rightarrow 0$ or $\eta\rightarrow 0.5$.

Figures 2(a), 3(a), 3(b), and 4(a) show cases with no staggered-ladder spectrum, i.e., with $g=1$ for all $\eta$. Figure 2(a) shows the Hofstadter butterfly for $\hbar _{\mathrm{S}}=2$ ($\hbar =4\pi$) and $\beta =0$, discovered in Ref. \cite{6}. For $\hbar _{\mathrm{S}}=l/2$ ($l$ odd) and $\beta$ with $g=1$ for all $\eta$, we find that the spectrum as function of $\eta$ looks like that illustrated in Figs. 3(a) and 3(b) for $l=1,3$: one has a chain of $l$ spectral structures, each resembling a ``curved" Hofstadter butterfly, and a ``reflection" of this chain. The spectral structure in Fig. 4(a), for $\hbar _{\mathrm{S}}=2/3$, also bears some resemblance to a Hofstadter butterfly.        
\begin{figure}[tbp]
\includegraphics[width=8.0cm]{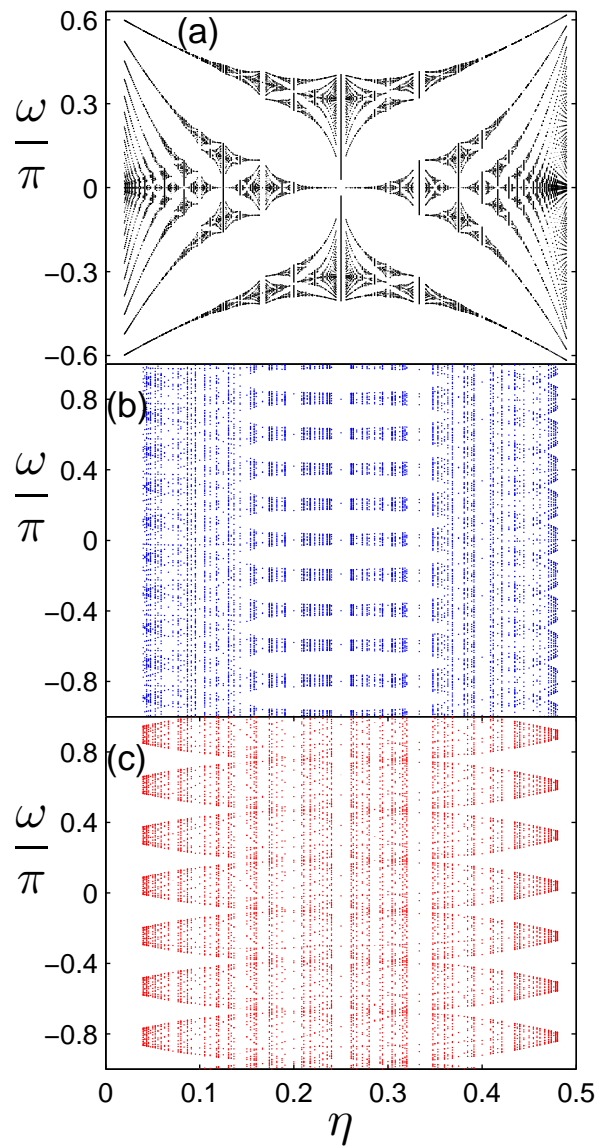}
\caption{(Color online) QE spectra $\omega $ of $\beta $-DKRs as functions of rational $\eta $ for $\hbar _{\mathrm{S}}=2/1$ ($\hbar =4\pi$), $K/\hbar =\tilde{K}/\hbar =1$, $V(\theta )=\tilde{V}(\theta )=\cos(\theta )$, and: (a) $\beta =0$, (b) $\beta =1/20$, (c) $\beta =1/7$. Cases (b) and (c) clearly feature a staggered-ladder structure with $g=10$ and $g=7$ stages, respectively, for most $\eta$'s; these values of $g$ coincide with the corresponding values of $g_{\rm max}$ in Eq. (\ref{gb}). In case (a), there is no staggered-ladder structure since $g=1$ for all $\eta$. The similarity of the spectral structure in this case to that of a Hofstadter butterfly was first pointed out in Ref. \cite{6}; see also Sec. V. In (a), (b), and (c), $\eta $ takes all rational values in $(0,0.5)$ with denominators $c\leq 50$, $c\leq 25$, and $c\leq 25$, respectively.}
\label{fig2}
\end{figure}
\begin{figure}[tbp]
\includegraphics[width=8.0cm]{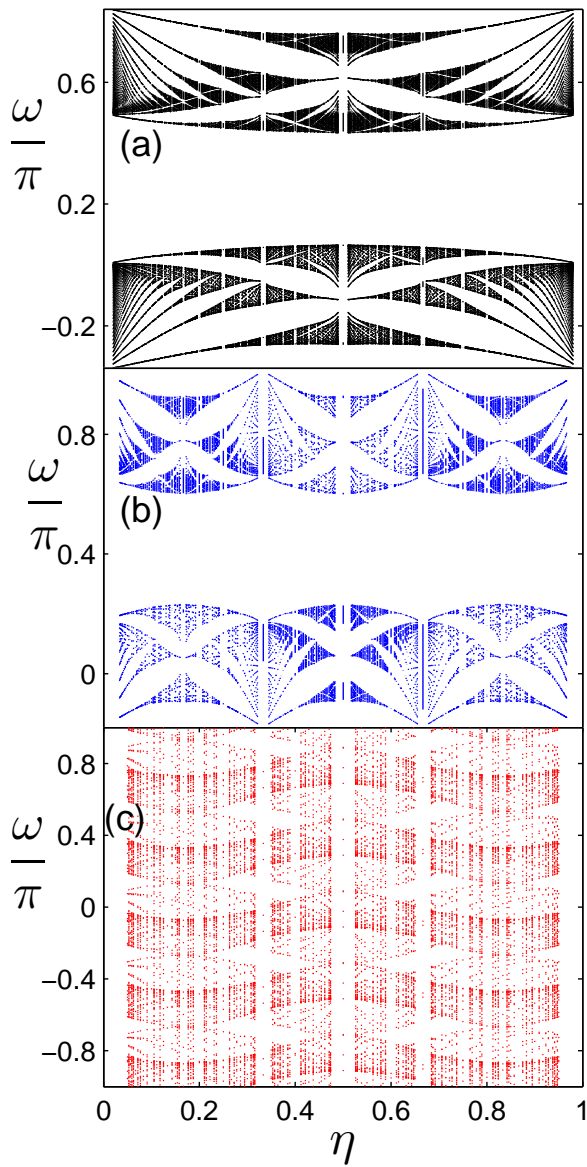}
\caption{(Color online) Similar to Fig. 2 but with the following changes: (a) $\hbar _{\mathrm{S}} =1/2$ and $\beta =0$; (b) $\hbar _{\mathrm{S}} =3/2$ and $\beta =1/3$; (c) $\hbar _{\mathrm{S}} =3/2$ and $\beta =1/5$. In cases (a) and (b), there appear structures resembling Hofstadter butterflies. Case (c) features a staggered-ladder structure with $g=g_{\rm max}=5$ stages for most values of $\eta$, where $g_{\rm max}$ is given by Eq. (\ref{gb}). This structure does not emerge in cases (a) and (b), since $g=1$ for all $\eta$. In (a), (b), and (c), $\eta$ takes all rational values in $(0,1)$ with denominators $c\leq 50$, $c\leq 30$, and $c\leq 20$, respectively.}  
\label{fig3}
\end{figure}
\begin{figure}[tbp]
\includegraphics[width=8.0cm]{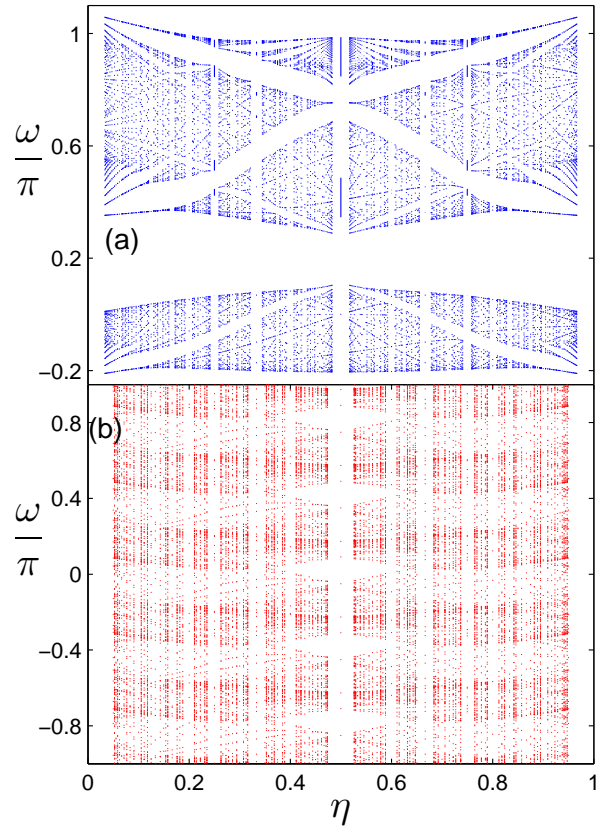}
\caption{(Color online). Similar to Fig. 2 but with the following changes: (a) $\hbar _{\mathrm{S}} =2/3$ and $\beta =0$; (b) $\hbar _{\mathrm{S}} =2/3$ and $\beta =1/5$. Only the latter case features a staggered-ladder structure with $g=g_{\rm max}=5$ stages for most values of $\eta$. The spectral structure in case (a), with $g=1$ for all $\eta$, bears some resemblance to a Hofstadter butterfly. In cases (a) and (b), $\eta$ takes all rational values in $(0,1)$ with denominators $c\leq 30$ and $c\leq 20$, respectively.}
\label{fig4}
\end{figure}

Finally, consider the case of irrational $\beta $, corresponding to the
limit of $g,\bar{r}\rightarrow \infty $ in Eq. (\ref{rb}). Assuming that the 
$\bar{q}=gc^{\prime }q$ bands are isolated (non-degenerate) in this limit, the width of each band must tend to zero. Then, the full QE spectrum will
consist of $c^{\prime }q$ infinite ladders of levels with corresponding
eigenstates that are localized in angular-momentum space (see also end of
Sec. IIIB).

\begin{center}
\textbf{IV. FREE QUANTUM DYNAMICS OF DKP}
\end{center}

We saw in Sec. IIIA that the quantum DKP at fixed quasimomentum $\beta \hbar 
$ reduces to a generalized DKR, the $\beta $-DKR. In this section, we
consider the quantum dynamics of the DKP for initial wave-packets $\varphi
_{\beta ,\omega }(x)$ associated with all the translated quasimomenta $\beta
^{(j)}\hbar $ [see Eq. (\ref{bj})] and the respective QEs (\ref{qej}). These
wave-packets are arbitrary linear combinations of the corresponding
eigenstates (\ref{qesj}):
\begin{equation}
\varphi _{\beta ,\omega }(x)=\sum_{j=-\infty }^{\infty }\zeta _{j}\exp
(i\beta ^{(j)}x)u_{\beta ^{(j)},\omega ^{(j)}}(x),  \label{iwp}
\end{equation}
where $\zeta _{j}$ are arbitrary but properly normalized coefficients, see
Eq. (\ref{nc}) below. After $t$ kicks, the wave-packet (\ref{iwp}) will
evolve to
\begin{equation}
\varphi _{\beta ,\omega ,t}(x)=\sum_{j=-\infty }^{\infty }\zeta _{j}\exp
(i\beta ^{(j)}x)u_{\beta ^{(j)},\omega ^{(j)}}(x)\exp (-i\omega ^{(j)}t).
\label{iwpt}
\end{equation}
Then, the expectation value of any physical observable in the wave-packet 
(\ref{iwpt}), as a function of the discrete time $t$, is given by a linear
combination of the factors $\exp [i(\omega ^{(j^{\prime })}-\omega ^{(j)})t]$
for all $j$ and $j^{\prime }$; see the explicit expression (\ref{ev}) below.
Now, from Eq. (\ref{qej}), the evolution frequencies
\begin{equation}
\omega ^{(j^{\prime })}-\omega ^{(j)}=2\pi c(j^{\prime }-j)\left[ \beta +\pi
c\left( j^{\prime }+j\right) /\hbar \right] \ \ \text{\textrm{mod}}(2\pi )
\label{ojjt}
\end{equation}
are completely independent of the system nonintegrability (the kicking
potentials) and are thus the same as those in the free ($K=\tilde{K}=0$)
case; in this sense, the quantum dynamics for the wave-packets (\ref{iwpt})
is a \emph{free} one. This quantum dynamics emerges as an interference
between the translated QE states (\ref{qesj}) in Eq. (\ref{iwpt}). Classical
entities analogous to these states, such as the translated orbits considered
in Sec. II (see Fig. 1), cannot, of course, exhibit this quantum
interference effect.

We now derive explicit expressions for (\ref{iwpt}) and for expectation
values in these wave-packets. Using Eq. (\ref{qesj}) in Eq. (\ref{iwpt}) and
expanding the $2\pi $-periodic function $u_{\beta ,\omega }(x)$ in a Fourier
series,
\begin{equation*}
u_{\beta ,\omega }(x)=\sum_{n=-\infty }^{\infty }u_{n}\frac{\exp (inx)}
{\sqrt{2\pi \hbar }},
\end{equation*}
we get
\begin{align}
\varphi _{\beta ,\omega ,t}(x)& =\sum_{j=-\infty }^{\infty }\sum_{n=-\infty
}^{\infty }\zeta _{j}u_{n}\exp (-i\omega ^{(j)}t)  \notag \\
& \times \frac{\exp \left[ i(\beta \hbar +2\pi cj+n\hbar )x/\hbar \right] }
{\sqrt{2\pi \hbar }}.  \label{iwpte}
\end{align}
We note that Eq. (\ref{iwpte}) is an expansion in normalized momentum states 
$\left\vert p=\beta \hbar +2\pi cj+n\hbar \right\rangle $ with coefficients $\zeta _{j}u_{n}\exp (-i\omega ^{(j)}t)$. The normalization condition for 
(\ref{iwpte}) is therefore:
\begin{equation}
\sum_{j=-\infty }^{\infty }\sum_{n=-\infty }^{\infty }|\zeta
_{j}u_{n}|^{2}=1.  \label{nc}
\end{equation}
Condition (\ref{nc}) can be satisfied if $\zeta _{j}$ and $u_{n}$ decay
sufficiently fast as $|j|,|n|\rightarrow \infty $. The coefficients $\zeta
_{j}$ can always be chosen to decay fast and such a decay for $u_{n}$ is
expected for irrational $\hbar _{\mathrm{S}}$ and/or $\beta $, see end of
Secs. IIIB and IIIE. We also note that the function (\ref{iwpte}), without
the factor $\exp (i\beta x)$, exhibits two periodicities in $x$, with basic
periods $2\pi $ and $\hbar /c$. For irrational $\hbar _{\mathrm{S}}$, these
periods are incommensurate and the wave-packet (\ref{iwpte}) is
quasiperiodic in $x$. For rational $\hbar _{\mathrm{S}}=l/q$ (and irrational 
$\beta $), $|\varphi _{\beta ,\omega ,t}(x)|$ is periodic with
period $2\pi l^{\prime }$, where $l^{\prime }$ is defined by $c/l=c^{\prime
}/l^{\prime }$ with $(c^{\prime },l^{\prime })$ coprime integers. If one keeps in Eq. (\ref{iwpt}) only terms with $j$ multiple of $l^{\prime }$, $j=sl^{\prime }$ ($s$ integer), all the connected quasimomenta (\ref{bj}) are equal, $\beta^{(sl^{\prime  })}=\beta$. Then, using Eq. (\ref{ubj}) with $n^{(sl^{\prime })}=sc^{\prime }q$ and Eq. (\ref{qerl}), Eq. (\ref{iwpt}) reduces to
\begin{equation}
\varphi _{\beta ,\omega ,t}(x)=e^{i(\beta x-\omega t)}u_{\beta ,\omega }(x)\sum_{s=-\infty }^{\infty }\zeta _{sl^{\prime }}\exp [isc^{\prime }q(x-\beta^{\prime }\hbar t)],
\label{tw}
\end{equation}
where $\beta^{\prime }=\beta+c^{\prime }q/2$. The sum in Eq. (\ref{tw}) is a traveling-wave function of $x-\beta^{\prime }\hbar t$, moving without change of shape at constant velocity $\beta^{\prime }\hbar$. As in the case of the ordinary kicked particle ($\eta =0$) \cite{d3}, this traveling-wave component of $\varphi _{\beta ,\omega ,t}(x)$ will be clearly exhibited by the quantum dynamics at least in some cases.    

For simplicity and without loss of generality, we assume physical
observables $\hat{O}$ (Hermitian operators) that are only functions of 
$\hat{x}$, $\hat{O}=O(\hat{x})$ \cite{note2}. Writing $O(x)=\int_{-\infty
}^{\infty }dp\bar{O}(p)\exp (ipx/\hbar )$, where $\bar{O}(p)$ is the Fourier
transform of $O(x)$, the expectation value of $\hat{O}$ in the wave-packet 
(\ref{iwpte}) is easily evaluated: 
\begin{align}
\left\langle \varphi _{\beta ,\omega ,t}\right\vert \hat{O}\left\vert
\varphi _{\beta ,\omega ,t}\right\rangle & =\sum_{j=-\infty }^{\infty
}\sum_{j^{\prime }=-\infty }^{\infty }\zeta _{j}\zeta _{j^{\prime }}^{\ast }
\bar{O}_{j,j^{\prime }}  \notag \\
& \times \exp \left[ i\left( \omega ^{(j^{\prime })}-\omega ^{(j)}\right) 
t\right] ,  \label{ev}
\end{align}
where
\begin{equation*}
\bar{O}_{j,j^{\prime }}=\sum_{n=-\infty }^{\infty }\sum_{n^{\prime }=-\infty
}^{\infty }u_{n}u_{n^{\prime }}^{\ast }\bar{O}\left[ 2\pi c\left( j^{\prime
}-j\right) +\left( n^{\prime }-n\right) \hbar \right] .
\end{equation*}

\begin{center}
\textbf{V. SUMMARY AND CONCLUSIONS}
\end{center}

In this paper, we have determined several quantum properties of double kicked systems (DKRs and DKPs) with classical translational invariance in momentum. This invariance exists for arbitrary rational values $a/c$ of $\eta$ in the Hamiltonian (\ref{HDKR}). Our results in Sec. III for $\beta$-DKRs, i.e., DKPs at fixed quasimomentum $\beta\hbar$, can be classified into three types. The first type of results hold for arbitrary values of $\beta$ and the scaled Planck constant $\hbar _{\mathrm{S}}$: 1) The QE eigenvalue problem of a $\beta$-DKR is equivalent to the energy eigenvalue problem of a finite strip of coupled lattice chains (\ref{mcs}) with translationally invariant hopping constants given by the simple expression (\ref{Wsn}). 2) A result following from 1) is the existence of exact connections between QE spectra of $\beta$-DKRs for $\beta$ in the set (\ref{bj}), which is infinite for irrational $\hbar _{\mathrm{S}}$; for rational $\hbar _{\mathrm{S}}=l/q$, this set contains $l^{\prime}$ elements, where $l^{\prime}$ is defined by $c/l=c^{\prime}/l^{\prime}$ with coprime $c^{\prime}$ and $l^{\prime}$.

The second type of results are valid for arbitrary $\beta$ but rational $\hbar _{\mathrm{S}}=l/q$ (Sec. IIIE): The QE spectrum of a $\beta$-DKR is a staggered ladder, a superposition of $c^{\prime}q$ ladder spectra. The latter spectra, all with the same spacing, are infinite or finite depending on whether $\beta$ is irrational or rational.

The third type of results are valid only for rational $\beta$ and $\hbar _{\mathrm{S}}$ (Secs. IIID and IIIE). Rational values of $\eta$, $\beta$, and $\hbar _{\mathrm{S}}$ have been shown to be the general necessary conditions for QR. This considerably generalizes the QR conditions for the DKR with $\beta =0$ and $\hbar =4\pi $, derived recently \cite{12}. The staggered-ladder QE spectrum is generally a superposition of $c^{\prime}q$ band ladders, each consisting of $g$ bands, where $g$ is the smallest integer such that $gc^{\prime }l(\beta+c^{\prime }q/2)$ is integer. 

In Sec. IV, we have considered DKP wave-packets that are superpositions of $\beta$-DKR QE states with connected spectra, i.e., with $\beta$ in the set (\ref{bj}). The time evolution of expectation values of observables in these wave-packets is independent of the nonintegrability, and the quantum dynamics is thus free. For rational $\hbar _{\mathrm{S}}=l/q$, the set (\ref{bj}) consists of $l^{\prime}$ elements and the wave-packet amplitude is periodic in $x$ with period $2\pi l^{\prime}$.

We thus see that all the results for rational $\hbar _{\mathrm{S}}$ exhibit number-theoretical features involving the coprime integers $c^{\prime}$ and $l^{\prime}$; the latter depend erratically on \emph{both} the rational value of  $\eta$ (specifying the classical translational invariance in momentum) and of $\hbar _{\mathrm{S}}$ (a purely quantum entity). If also $\beta$ is rational (QR case), one has to consider another integer $g$, depending erratically on all quantities $\eta$, $\hbar _{\mathrm{S}}$, and $\beta$.

In the case of the ordinary kicked particle ($\eta=0$), the staggered-ladder QE spectra for rational $\hbar _{\mathrm{S}}$ were shown \cite{d3} to have several quantum-dynamical manifestations: A suppression of quantum resonances for rational $\beta$ and a dynamical localization for irrational $\beta$ characterized by unique features such as traveling-wave components in the time evolution. The staggered-ladder QE spectra for arbitrary rational values of $\eta$ and $\hbar _{\mathrm{S}}$ will lead to similar phenomena, depending now on the number-theoretical quantities $c^{\prime}$ and $l^{\prime}$; see, for example, the traveling-wave component given by the sum in Eq. (\ref{tw}). These phenomena, which are robust under small variations of $\beta$ \cite{d3}, should be experimentally realizable by atom-optics methods for at least low-order rational values of $\eta$ and $\hbar _{\mathrm{S}}$, as it was shown in the $\eta=0$ case \cite{eqr,qam,shsn,drrts}.

For $g=1$, i.e., in the absence of spectral ladders, $\beta$-DKRs appear to exhibit a variety of interesting spectral structures for different values of $\hbar$, as shown in Figs. 2(a), 3(a), 3(b), and 4(a). The structure in Fig. 2(a) for $\hbar =4\pi$ and $\beta =0$, corresponding to the \textquotedblleft on-resonance\textquotedblright\ DKR \cite{6,7,8,9,10,11,12,13}, is the approximate Hofstadter butterfly discovered in Ref. \cite{6}. This structure can be understood from the fact, recently shown \cite{13}, that a one-parameter family of on-resonance DKRs for rational $\eta$ is unitarily equivalent to an analogous family of kicked Harper models \cite{khm,gkp,gkpr,d4,d5,dfw}, which are known \cite{gkp} to feature a Hofstadter-butterfly structure. For irrational $\eta$, already the QE spectrum of a single on-resonance DKR exactly coincides with that of a kicked Harper model \cite{9}.

Understanding new spectral structures for $\hbar\neq 4\pi$, such as those in Figs. 3(a), 3(b), and 4(a) which resemble again Hofstadter butterflies, is a problem that we plan to consider in future works. The sensitivity of these structures and that in Fig. 2(a) to the value of $\beta$ is mainly due to the occurrence of spectral ladders for generic $\beta$. Since these ladders are entirely a consequence of classical translational invariance in momentum for rational $\eta$, we expect that the spectral structures will not be so sensitive to $\beta$ for irrational $\eta$. A systematic study to confirm this expectation is also planned for future works.
         
%
\begin{center}
\textbf{APPENDIX A}
\end{center}

We derive here the relations (\ref{ceta})-(\ref{cb}) from the commutation
relation $\left[ \hat{D}_{\bar{q}\hbar },\hat{U}(\theta ,\hat{L}+\beta \hbar
)\right] =0$, where $\hat{D}_{\bar{q}\hbar }=\exp (-\bar{q}\theta )$ 
($\bar{q}$ integer). For simplicity, we denote here  
$\hat{U}(\theta ,\hat{L}+\beta \hbar )$ by $\hat{U}_{\beta }$. From Eq. (\ref{U}) we have 
\begin{align}
\hat{U}_{\beta }& =e^{-i\left( 1-\eta \right) (\hat{L}+\beta \hbar
)^{2}/(2\hbar )}e^{-i\tilde{K}\tilde{V}(\theta )/\hbar }  \notag \\
& \times e^{-i\eta (\hat{L}+\beta \hbar )^{2}/(2\hbar )}e^{-iKV(\theta
)/\hbar }.  \label{Ub}
\end{align}
Using the fact that $\hat{D}_{\bar{q}\hbar }=\exp (-\bar{q}\theta )$ is a
translation of $\hat{L}$ by $\bar{q}\hbar $ while $\exp (i\gamma \hat{L})=\exp (\gamma \hbar d/d\theta )$ (arbitrary $\gamma $) is a translation of 
$\theta $ by $\gamma \hbar $, we easily find that the application of 
$\hat{D}_{\bar{q}\hbar }$ on (\ref{Ub}) yields:
\begin{align}
\hat{D}_{\bar{q}\hbar }\hat{U}_{\beta }& =e^{-i\left( 1-\eta \right) (\hat{L}+\beta \hbar )^{2}/(2\hbar )}e^{-i\tilde{K}\tilde{V}[\theta -(1-\eta )\bar{q}\hbar ]/\hbar }  \notag \\
& \times e^{-i\eta (\hat{L}+\beta \hbar )^{2}/(2\hbar )}e^{-iKV(\theta -
\bar{q}\hbar )/\hbar }  \notag \\
& \times e^{-i\bar{q}\hat{L}}e^{-i(\beta \bar{q}\hbar +\bar{q}^{2}\hbar /2)}
\hat{D}_{\bar{q}\hbar }.  \label{DUb}
\end{align}
To get $[ \hat{D}_{\bar{q}\hbar },\hat{U}_{\beta }] =0$ for general potentials $V(\theta )$ and $\tilde{V}(\theta )$, the first two lines of Eq. (\ref{DUb}) must be identified with $\hat{U}_{\beta }$ and the product of the first two factors in the third line must be identically $1$. The first two lines give $\hat{U}_{\beta }$ 
only if $\bar{q}\hbar =2\pi \bar{c}$ and $(1-\eta )\bar{c}=d$, where $\bar{c}$ and $d$ are integers. The latter equations are Eqs. (\ref{ceta}) and 
(\ref{hbar}). The first factor in the third line of Eq. (\ref{DUb})  is
identically $1$ since $\exp (-i\bar{q}\hat{L})=\exp (-\bar{q}\hbar d/d\theta
)=\exp (-2\pi \bar{c}d/d\theta )$ is a translation of $\theta $ by $-2\pi 
\bar{c}$. Finally, the second factor is equal to $1$ only if Eq. 
(\ref{cb}) is satisfied.

\begin{center}
\textbf{APPENDIX B}
\end{center} 

We show here that under QR conditions the QE spectrum of a $\beta $-DKR generally consists of $\bar{q}$ bands, where $\bar{q}$ is given by Eq. (\ref{qb}). We first derive a general expression for the QE states of the $\beta $-DKR. These states are simultaneous eigenstates of the commuting operators $\hat{D}_{\bar{q}\hbar }$ and $\hat{U}(\theta ,\hat{L}+\beta \hbar )$. Since $\hat{D}_{\bar{q}\hbar }=\exp (-i\bar{q} \theta )$, the functions $\sum_{n=-\infty }^{\infty }\delta \lbrack \theta
-2\pi n-2\pi m/\bar{q}-\alpha ]$, for $m=0,\dots ,\bar{q}-1$ and
\textquotedblleft quasiangle\textquotedblright\ $\alpha $ varying in the
Brillouin zone (BZ) $0\leq \alpha <2\pi /\bar{q}$, clearly form a complete
set of eigenstates of $\hat{D}_{\bar{q}\hbar }$ with eigenvalues $\exp (-i
\bar{q}\alpha )$ independent of $m$. Therefore, the simultaneous eigenstates
of $\hat{D}_{\bar{q}\hbar }$ and $\hat{U}(\theta ,\hat{L}+\beta \hbar )$
will be generally given by $\bar{q}$ independent linear combinations of
these functions over $m$,
\begin{align}
u_{b,\alpha ,\beta }(\theta )& =\sum_{m=0}^{\bar{q}-1}\phi _{b}(m;\alpha
,\beta )  \notag \\
& \times \sum_{n=-\infty }^{\infty }\delta \left( \theta -2\pi n-\frac{2\pi m
}{\bar{q}}-\alpha \right) ,  \label{seis}
\end{align}
$b=1,\dots ,\bar{q}$. Then, when $\alpha $ varies in the BZ $[0,2\pi /\bar{q})$,
the QEs corresponding to (\ref{seis}) form $\bar{q}$ bands $\omega
_{b}(\alpha ,\beta )$ labeled by $b$. Since $\bar{q}=gc^{\prime }q$ is
minimal for the given rational value of $\beta $, the BZ is maximal, so that
these $\bar{q}$ bands can be expected to be generally isolated (non-degenerate).

It is instructive to see how the $\bar{q}$ bands can be determined, at least in principle, from the coupled lattice chains (\ref{mcs}) for the $\beta $-DKR. Let us first show that these chains are essentially periodic in $n$ with period $\bar{q}$. If $\bar{r}$ is the integer in the QR condition $gc^{\prime }l(\beta+gc^{\prime }q/2)=\bar{r}$, let us define $u^{\prime}_{k,n}=\exp [2\pi i\bar{r}kn/(c\bar{q})]u_{k,n}$, $W^{\prime}_{k,n-r}=\exp [2\pi i\bar{r}k(n-r)/(c\bar{q})]W_{k,n-r}$, and the Fourier transforms $\bar{u}^{\prime}_{s,n}$ and $\bar{W}^{\prime}_{s,n}$ as in Eqs. (\ref{usn}) and (\ref{Wsn}). We then easily find that Eqs. (\ref{mcs}) hold with $\bar{u}$ and $\bar{W}$ replaced by $\bar{u}^{\prime}$ and $\bar{W}^{\prime}$, respectively, and with $\omega$ in Eq. (\ref{Tsn}) replaced by $\omega +2\pi\bar{r}n/\bar{q}$. Using the QR conditions and $\bar{q}=gc^{\prime}q$, the modified Eqs. (\ref{mcs}) can be seen to be invariant under the simultaneous transformations $n\rightarrow n+\bar{q}$ and $r\rightarrow r+\bar{q}$. This implies, by Bloch theorem, that $\bar{u}^{\prime}_{s,n}$ can be chosen to be equal to $\bar{u}^{\prime}_{s,n+\bar{q}}$ up to a phase factor. Using the latter boundary condition, the modified Eqs. (\ref{mcs}) reduce to the eigenvalue problem of a $c\bar{q}\times c\bar{q}$ matrix depending on a phase. The solution of this problem, as explained in Sec. IIIB, yields $c$ replicas of the $\bar{q}$ QE bands.


\begin{thebibliography}{99}
\bibitem{bc} B.V. Chirikov, Phys. Rep. \textbf{52}, 263 (1979).

\bibitem{cm} J.R. Cary and J.D. Meiss, Phys. Rev. A \textbf{24}, 2664 (1981).

\bibitem{krw} C.F.F. Karney, A.B. Rechester, and R.B. White, Physica D 
\textbf{4}, 425 (1982).

\bibitem{ikh} Y.H. Ichikawa, T. Kamimura, and T. Hatori, Physica D 
\textbf{29}, 247 (1987).

\bibitem{ihkm} R. Ishizaki, T. Horita, T. Kobayashi, and H. Mori, Prog.
Theor. Phys. \textbf{85}, 1013 (1991).

\bibitem{d} I. Dana, Phys. Rev. E \textbf{69}, 016212 (2004).

\bibitem{d1} O. Barash and I. Dana, Phys. Rev. E \textbf{71}, 036222 2005; 
\textbf{75}, 056209 (2007).

\bibitem{wm} G.M. Zaslavskii, M.Yu. Zakharov, R.Z. Sagdeev, D.A. Usikov, and
A.A. Chernikov, Sov. Phys. JETP \textbf{64}, 294 (1986); I. Dana and M.
Amit, Phys. Rev. E \textbf{51}, R2731 (1995).

\bibitem{wm1} I. Dana and T. Horesh, Lect. Notes Phys. \textbf{511}, 51
(1998).

\bibitem{gmz} G.M. Zaslavsky, Phys. Rep. \textbf{371}, 461 (2002), and
references therein.

\bibitem{cr1} H. Schanz, M.-F. Otto, R. Ketzmerick, and T. Dittrich, Phys.
Rev. Lett. \textbf{87}, 070601 (2001); H. Schanz, T. Dittrich, and R.
Ketzmerick, Phys. Rev. E \textbf{71}, 026228 (2005).

\bibitem{cr2} T. Cheon, P. Exner, and P. Seba, J. Phys. Soc. Jpn. 
\textbf{72}, 1087 (2003).

\bibitem{cr3} L. Cavallasca, R. Artuso, and G. Casati, Phys. Rev. E 
\textbf{75}, 066213 (2007).

\bibitem{cr4} I. Dana and V.B. Roitberg, Phys. Rev. E \textbf{83}, 066213
(2011).

\bibitem{dr} I. Dana and W.P. Reinhardt, Physica D \textbf{28}, 115 (1987).

\bibitem{dd} I. Dana, N.W. Murray, and I.C. Percival, Phys. Rev. Lett. 
\textbf{62}, 233 (1989); I. Dana, Physica D \textbf{39}, 205 (1989); O.
Barash and I. Dana, Phys. Rev. E \textbf{74}, 056202 (2006).

\bibitem{gpf} D. R. Grempel, R. E. Prange, and S. Fishman, Phys. Rev. A 
\textbf{29}, 1639 (1984), and references therein.

\bibitem{fmi} F.M. Izrailev, Phys. Rep. \textbf{196}, 299 (1990), and
references therein.

\bibitem{fgr} S. Fishman, I. Guarneri, and L. Rebuzzini, J. Stat. Phys. 
\textbf{110}, 911 (2003).

\bibitem{wgf} S. Wimberger, I. Guarneri, and S. Fishman, Nonlinearity 
\textbf{16}, 1381 (2003).

\bibitem{d2} I. Dana and D.L. Dorofeev, Phys. Rev. E \textbf{73}, 026206
(2006).

\bibitem{d3} I. Dana, Phys. Rev. A \textbf{87}, 043623 (2013).

\bibitem{khm} P. Leboeuf \textit{et al.}, Phys. Rev. Lett. \textbf{65}, 3076
(1990).

\bibitem{gkp} T. Geisel, R. Ketzmerick, and G. Petschel, Phys. Rev. Lett. 
\textbf{67}, 3635 (1991).

\bibitem{gkpr} R. Artuso \textit{et al.}, Int. J. Mod. Phys. B \textbf{8}, 207 (1994), and references therein.

\bibitem{d4} I. Dana, Phys. Rev. Lett. \textbf{73}, 1609 (1994).

\bibitem{d5} I. Dana, Phys. Rev. E \textbf{52}, 466 (1995),

\bibitem{dfw} I. Dana, M. Feingold, and M. Wilkinson, Phys. Rev. Lett. 
\textbf{81}, 3124 (1998); I. Dana, Y. Rutman, and M. Feingold, Phys. Rev. E 
\textbf{58}, 5655 (1998).

\bibitem{dd1} I. Dana and D.L. Dorofeev, Phys. Rev. E \textbf{72}, 046205
(2005).

\bibitem{lw} E. Lundh and M. Wallin, Phys. Rev. Lett. \textbf{94}, 110603
(2005). 

\bibitem{d6} I. Dana and V. Roitberg, Phys. Rev. E \textbf{76}, 015201(R)
(2007).

\bibitem{kgp} A. Kenfack, J. Gong, and A. K. Pattanayak, Phys. Rev. Lett. 
\textbf{100}, 044104 (2008).

\bibitem{d7} I. Dana, Phys. Rev. E \textbf{81}, 036210 (2010); J. Phys.:
Conf. Ser. \textbf{285}, 012048 (2011).

\bibitem{eqr} C. Ryu \textit{et al.}, Phys. Rev. Lett. \textbf{96}, 160403 (2006).

\bibitem{qam} M.K. Oberthaler, R.M. Godun, M.B. d'Arcy, G.S. Summy, and K.
Burnett, Phys. Rev. Lett. \textbf{83}, 4447 (1999); R.M. Godun, M.B. d'Arcy,
M.K. Oberthaler, G.S. Summy, and K. Burnett, Phys. Rev. A \textbf{62},
013411 (2000); M.B. d'Arcy, R.M. Godun, M.K. Oberthaler, G.S. Summy, K.
Burnett, and S.A. Gardiner, Phys. Rev. E \textbf{64}, 056233 (2001); S.
Schlunk, M.B. d'Arcy, S.A. Gardiner, D. Cassettari, R.M. Godun, and G.S.
Summy, Phys. Rev. Lett. \textbf{90}, 054101 (2003); S. Schlunk, M.B. d'Arcy,
S.A. Gardiner, and G.S. Summy, \textit{ibid.} \textbf{90}, 124102 (2003);
Z.-Y. Ma, M.B. d'Arcy, and S.A. Gardiner, \textit{ibid.} \textbf{93}, 164101
(2004); G. Behinaein, V. Ramareddy, P. Ahmadi, and G.S. Summy, \textit{ibid.}
\textbf{97}, 244101 (2006).

\bibitem{shsn} M. Sadgrove, M. Horikoshi, T. Sekimura, and K. Nakagawa,
Phys. Rev. Lett. \textbf{99}, 043002 (2007).

\bibitem{drrts} I. Dana, V. Ramareddy, I. Talukdar, and G. S. Summy, Phys.
Rev. Lett. \textbf{100}, 024103 (2008); I. Dana, V. B. Roitberg, V.
Ramareddy, I. Talukdar, and G. S. Summy, Int. J. Bifurcation Chaos 
\textbf{20}, 255 (2010).

\bibitem{note} The web map $M_{\mathrm{w}}$ is the Poincar\'{e} map of a
periodically kicked harmonic oscillator (or, equivalently, a periodically kicked charge in a uniform magnetic field) \cite{wm}. Let the time period of the kicking be $T$ and the frequency of the oscillator be $\nu$. Then, only for rational values of $\nu T$ with denominator $n=1,2,3,4,6$ the map $M_{\mathrm{w}}^{n}$ exhibits translational invariance in phase space. As a consequence, one gets, for $n=3,4,6$, a global chaotic motion on an infinite stochastic web for arbitrarily small kicking strength; for $n=1,2$, the system is integrable. 

\bibitem{des} I. Dana, E. Eisenberg, and N. Shnerb, Phys. Rev. E 
\textbf{54}, 5948 (1996).

\bibitem{1} T. Jonckheere \textit{et al.}, Phys. Rev. Lett. \textbf{91},
253003 (2003).

\bibitem{2} P.H. Jones \textit{et al.}, Phys. Rev. Lett. \textbf{93}, 223002
(2004).

\bibitem{3} G. Hur \textit{et al.}, Phys. Rev. A \textbf{72}, 013403 (2005).

\bibitem{4} J. Wang \textit{et al.}, Phys. Rev. Lett. \textbf{99}, 234101
(2007).

\bibitem{5} J. Gong and J. Wang, Phys. Rev. E \textbf{76}, 036217 (2007).

\bibitem{6} J. Wang and J. Gong, Phys. Rev. A \textbf{77}, 031405(R) (2008).

\bibitem{7} J. Wang and J. Gong, Phys. Rev. E \textbf{78}, 036219 (2008).

\bibitem{8} J. Wang, A.S. Mouritzen, and J.B. Gong, J. Mod. Optics 
\textbf{56}, 722 (2009).

\bibitem{9} W. Lawton, A.S. Mouritzen, J. Wang, and J. Gong, J. Math. Phys. 
\textbf{50}, 032103 (2009).

\bibitem{10} J. Wang, I. Guarneri, G. Casati, and J.B. Gong, Phys. Rev.
Lett. \textbf{107}, 234104 (2011).

\bibitem{11} D.Y.H. Ho and J.B. Gong, Phys. Rev. Lett. \textbf{109}, 010601
(2012).

\bibitem{12} H.L. Wang, J. Wang, I. Guarneri, G. Casati, and J.B. Gong,
Phys. Rev. E \textbf{88}, 052919 (2013).

\bibitem{13} H.L. Wang, D.Y.H. Ho, W. Lawton, J. Wang, and J.B. Gong, Phys.
Rev. E \textbf{88}, 052920 (2013).

\bibitem{hb} D.R. Hofstadter, Phys. Rev. B \textbf{14}, 2239 (1976).

\bibitem{note1} Actually, $gc^{\prime }l(\beta +c^{\prime }q/2)=\bar{r}$ gives 
$\beta =\bar{r}/(gc^{\prime }l)-c^{\prime }q/2\ \ $\textrm{mod}$(1)$, which is the same as Eq. (\ref{rb}) if we replace $\bar{r}$ by $\bar{r}-(g^{2}-g)c^{\prime 2}ql/2$.

\bibitem{note2} In the extremely opposite case of observables depending only
on $\hat{p}$, one can easily see that the expectation values are time
independent.

\end{thebibliography}
\end{document}